\documentclass[%
 reprint,
 amsmath,amssymb,
 aps,
 pra,
 onecolumn
]{revtex4-2}
\usepackage{float} 
\usepackage{graphicx}
\usepackage{dcolumn}
\usepackage{bm}
\usepackage{hyperref}
\usepackage{xcolor}
\usepackage{setspace}
\definecolor{Fire}{RGB}{234,92,12}
\definecolor{Water}{RGB}{95, 130, 193}
\definecolor{Sky}{RGB}{152, 197, 221}
\definecolor{Purple}{RGB}{103, 82, 158}
\hypersetup{
    colorlinks=true,
    linkcolor=Water,
    citecolor=Water,
    urlcolor=Purple,
    pdftitle={Quantum statistical mechanics on a digital quantum computer},
    pdfkeywords={Trapped Ions} {DQC1} {Kernel Polynomial Method} {Quantum Simulation} {Density of States},
    pdfnewwindow=true,
    }

\usepackage{xcolor}
\makeatletter
\newsavebox{\@brx}
\newcommand{\llangle}[1][]{\savebox{\@brx}{\(\m@th{#1\langle}\)}%
  \mathopen{\copy\@brx\kern-0.5\wd\@brx\usebox{\@brx}}}
\newcommand{\rrangle}[1][]{\savebox{\@brx}{\(\m@th{#1\rangle}\)}%
  \mathclose{\copy\@brx\kern-0.5\wd\@brx\usebox{\@brx}}}
\makeatother

\usepackage{float}
\usepackage{dsfont}     
\usepackage{mathtools}
\usepackage{braket}
\usepackage{qcircuit}

\usepackage{xspace} 
\usepackage{multirow} 

\newcommand*{\figref}[2][]{\hyperref[{fig:#2}]{Fig.~\ref*{fig:#2}\ifx\\#1\\\else.($#1$)\fi}}

\newcommand*{\appref}[2][]{\hyperref[{app:#2}]{Appendix~\ref*{app:#2}}}

\renewcommand*{\eqref}[2][]{\hyperref[{#2}]{Eq.~(\ref*{#2})}}

\newcommand*{\secref}[2][]{\hyperref[{#2}]{Section~\ref*{#2}}}

\usepackage{ulem}

\begin{document}

\title{A powered full quantum eigensolver for energy band structures}
\
\author{Bozhi Wang}
\affiliation{State Key Laboratory of Low-Dimensional Quantum Physics and Department of Physics, Tsinghua University, Beijing 100084, China}
\author{Jingwei Wen}
\affiliation{China Mobile (Suzhou) Software Technology Company Limited, Suzhou 215163, China}
\author{Jiawei Wu}
\affiliation{Centre for Quantum Technologies, National University of Singapore, 119007, Singapore}
\author{Haonan Xie}
\author{Fan Yang}
\affiliation{State Key Laboratory of Low-Dimensional Quantum Physics and Department of Physics, Tsinghua University, Beijing 100084, China}
\affiliation{Beijing Academy of Quantum Information Sciences,  Beijing 100193, China}
\author{Shijie Wei}
\email{weisj@baqis.ac.cn}
\affiliation{Beijing Academy of Quantum Information Sciences,  Beijing 100193, China}
\author{Gui-lu Long}
\email{gllong@tsinghua.edu.cn}
\affiliation{State Key Laboratory of Low-Dimensional Quantum Physics and Department of Physics, Tsinghua University, Beijing 100084, China}
\affiliation{Beijing Academy of Quantum Information Science, Beijing 100193, China}
\affiliation{Frontier Science Center for Quantum Information, Beijing 100084, China}
\affiliation{Beijing National Research Center for Information Science and Technology, Beijing 100084, China}


\begin{abstract}

    There has been an increasing research focus on quantum algorithms for condensed matter systems recently, with a particular emphasis on calculating energy band structures. Here, we propose a quantum algorithm, the powered full quantum eigensolver(P-FQE), by using the exponentiation of operators of the full quantum eigensolver(FQE). This leads to an exponential increase in the success probability of measuring the target state in certain circumstances that the number of generating elements involved in the exponentiation of operators exhibit a logpolynomial dependence on the number of orbitals. Furthermore, we conduct numerical calculations for band structure determination of the twisted double-layer graphene. We experimentally demonstrate the feasibility and robustness of the P-FQE algorithm using  superconducting quantum computers for graphene and Weyl semimetal. One significant advantage of our algorithm is its ability to reduce the requirements of extremely high-performance hardware, making it more suitable for energy spectra determination on noisy intermediate-scale quantum (NISQ) devices.

\end{abstract}

\maketitle

\section{Introduction}

     Feynman\cite{feynman2018simulating} pointed that when classical computers were used to simulate quantum  systems, the resource consumption would exponentially increase. Nevertheless, by developing computing systems that work according to the laws of quantum mechanics, this difficulty can be  avoided in principle. Quantum computation has gained considerable attention across various fields \cite{benioff1980computer,montanaro2016quantum,bharti2022noisy,shor1999polynomial,lloyd1996universal,harrow2009quantum,yan2022factoring,wang2022variational,wei2023quantum}, harnessing its inherent quantum advantages. Quantum chemistry has emerged as one of the extensively studied applications, predominantly focused on atomic and molecular systems \cite{dumitrescu2018cloud,kandala2017hardware,aspuru2005simulated,nam2020ground,lv2023qcsh}. However, research efforts pertaining to more complex many-body systems like solid state are comparatively limited.

In solid-state physics, the concept of energy band is fundamental. It describes the distribution of energy levels available to electrons as a function of wavevector  within the material \cite{bloch1928quantum,wilson1931theory}. Calculating energy band structures holds paramount significance in understanding the electronic properties of materials, designing novel materials, and interpreting material behaviors.

The prevailing quantum algorithms for calculating the electronic band structure use hybrid quantum-classical methods like variational quantum eigensolver (VQE) \cite{peruzzo2014variational,cerezo2021variational,cerasoli2020quantum,yoshioka2022variational} for ground-state and its modified versions for excited-state, such as variational quantum deflation (VQD) \cite{jones2019variational,higgott2019variational,wen2021variational} and subspace-search variational quantum eigensolver (SSVQE) \cite{nakanishi2019subspace,parrish2019quantum}. These variational quantum algorithms (VQA) are studied vigorously for their low-depth circuits \cite{preskill2018quantum,bharti2022noisy}. It consists of parameterized quantum circuits and classical optimizations that optimize the expectation values measured from the quantum circuit to iterate parameters in the quantum circuit. The parameterized quantum circuit can be designed based on problem-inspired ansatze or hardware-efficient ansatze \cite{bharti2022noisy}. 
However, there are still some open questions concerning VQA, such as the existence of barren plateaus \cite{mcclean2018barren} and the ambiguity surrounding quantum advantage \cite{bittel2021training,barak2021classical}.

The FQE algorithm was proposed for quantum chemistry simulations \cite{wei2020full}, which simulates the objective Hamiltonian and performs gradient descent optimization entirely on the quantum computer. The parameters of quantum circuits in FQE remain fixed throughout the iteration process and are determined by the objective Hamiltonian. With each iteration of the quantum gradient descent circuit, the state vector becomes progressively closer to the ground state. Moreover, the extended version of FQE for excited-state named full quantum excited-state solver (FQESS) was proposed subsequently \cite{wen2021full}. By utilizing the measurement results of the lower energy levels obtained from the quantum circuit, the Hamiltonian and the parameters in the circuit can be updated to target the next energy level. The ground state of the updated Hamiltonian corresponds to the eigenvector associated with the next energy level. However, the main drawback of these algorithms is that the success probability decreases significantly with each iteration, which limits their applicability on NISQ devices. To address this issue, we propose an algorithm here named Powered-FQE, which directly implements multiple powers of the operator instead of executing multiple iterations of the operator on the quantum computer. This approach can significantly reduce the actual number of real runs, and in the extreme cases, it can achieve convergence with just a single run. This paper is organized as follows: In Sec. \ref{sec:method}, we present the framework of P-FQE. We then apply P-FQE to calculate the energy band structures of twisted double-layer graphene and compare the numerical simulation results with classical methods in Sec. \ref{sec:numericalsimulation}. In Sec. \ref{sec:results}, we conduct experiments on two superconducting quantum computers (Quafu and IBM quantum cloud platform) to compare and validate the effectiveness and robustness of our algorithm. Finally, we provide a conclusion in Sec. \ref{chap-con}.
\section{Method}
\label{sec:method}

\subsection{Band structure Hamiltonian}

The construction of energy band theory relies on two basic approximations. The first one is the Born-Oppenheimer approximation, which assumes that the electronic particles have a significantly smaller mass compared to atomic nuclei. Consequently, it is reasonable to neglect the motion of nuclei when studying electronic motion. This allows for the decomposition of the system wave function into a nuclear component and an electronic component. The second one is the single electron approximation, which simplifies complex multi-electron systems by replacing electron interactions with an average potential. The collective interactions between electrons and ion cores, as well as electron-electron interactions, can be effectively described by a single-electron potential field that possesses the same translational symmetry as the crystal lattice: $V(\vec{r}+\vec{R_l})=V(\vec{r})$, where $\vec{R_l}$ is lattice vector. The wavefunctions in periodic structures satisfy Bloch theorem: $\psi(\vec{r}+\vec{R_l}) = e^{i\vec{k} \cdot \vec{R_l}} \psi(\vec{r}) $, where $\vec{k}$ is wave vector. The single-electron Hamiltonian in a crystal can be expressed in the following form
\begin{equation}
    h(\vec{r}) = -\frac{\hbar^2}{2m}\nabla^2+V(\vec{r}).
\end{equation}

In the tight-binding (TB) approximation, only hopping terms among nearest neighbor sites need to be considered \cite{andersen1984explicit,goringe1997tight}. The total energy including both the interaction energy and kinetic energy can be unified as the hopping term, while the potential energy term or on-site energy simply adds a constant. In the second quantization formalism, the TB Hamiltonian of a crystal can be written as follows:
\begin{equation}
    H = \sum_{\left \langle i,j \right \rangle,m,n} t_{ij,mn}c_{i,m}^{\dagger}c_{j,n},
\end{equation}
where $m$, $n$ represent atomic orbitals, $\left \langle i,j \right \rangle$ denote pairs of nearest neighbor lattice. $c_{i,m}^{\dagger}(c_{i,m})$ are the creation (annihilation) operators acting at the orbital $m$ and site $i$. The $t_{ij,mn}$ denotes the hopping parameter between corresponding orbitals and sites, and it can be calculated using Wannier functions $\left \{ R_{mi} \right \}$ as the basis
\begin{equation}
    t_{ij,mn}=\left \langle R_{mi}|h(\vec{r}) |R_{nj}\right \rangle.
\end{equation}
Based on the TB approximation, the Wannier functions in the equation can be substituted with atomic orbital functions. Performing a Fourier transform of the creation and annihilation operators in the Wannier representation leads to the TB Hamiltonian ($H_k$) in the Bloch representation. In practice, the Hamiltonian in second quantization form is typically mapped into a qubit Hamiltonian using the Jordan-Wigner or Bravyi-Kitaev transformations \cite{seeley2012bravyi,batista2001generalized}
\begin{equation}
    H_k = \sum_{i} \alpha _{ik} P_i,
    \label{Hlcu}
\end{equation}
where $P_i= \sigma_{i1} \otimes \sigma_{i2}\otimes \cdots$ is a tensor product of Pauli matrices defined as ``Pauli word" with $\sigma \in \left \{ I, X, Y, Z \right \}$.

\subsection{The quantum algorithm}

\subsubsection{Power iteration method}
\label{sec:iteration method}

Power iteration method for computing eigenvalues and eigenvectors is valuable for estimating eigenvectors of large and sparse matrices \cite{panju2011iterative}.
Assuming matrix $A$ possesses $n$ eigenstates $u_i$ and the corresponding eigenvalues $\lambda_i$, the eigenfunctions can be expressed as $Au_i = \lambda_iu_i$.
Due to the completeness of the eigenfunctions of the Hermitian operator, arbitrary initial state of this quantum system can be fully described within the Hilbert space spanned by this set of eigenfunctions. This description involves representing the state as a linear superposition of eigenstates: $x^{(0)}=\sum_{i=0}^{n-1} a_iu_i$. Applying the $k$ power of $A$ to $x^{(0)}$ yields

\begin{equation}
    x^{(k)}=A^kx^{(0)}=\lambda _0^k[a_0u_0+\sum_{i=1}^{n-1}a_i(\frac{\lambda _i}{\lambda _0})^k u_i].
\end{equation}
Suppose that the absolute values of the eigenvalues satisfy the inequality constraints $|\lambda_0| \ge |\lambda_1| \ge |\lambda_2|\cdot \cdot \cdot  \ge |\lambda_n|$, we have $lim_{k \to \infty} (\lambda_i/\lambda_0)^k = 0$. So if the exponent $k$ is sufficiently large, $x^{(k)}$ will approach the ground state $u_0$. This method can also be interpreted from the perspective of quantum gradient descent, as shown in Ref \cite{wen2021full}. And we develop our quantum algorithm based on above process.

In FQE and FQESS algorithms, each iteration is essentially the evolution of quantum states under the same operator, but their probability of success decreases exponentially with the number of iterations. Our algorithm is based on the idea of reducing the number of iterations to improve the success probability, achieved by replacing the original evolution operator with the exponentiation of the evolution operator. We analyze the order of power required and estimate error in Appendix \ref{error estimation}. In our method, there exists a trade-off between the number of required iterations and the number of the powers. For simplicity, we present our algorithm by employing a single run.

\subsubsection{Vector encoding method}

In order to realize our algorithm on a quantum circuit, we need to decompose the evolution operator into a linear combination of unitary operators \cite{gui2006general,gui2008duality,gui2009allowable,childs2012hamiltonian,wen2019experimental,wen2020observation,wen2021stable}. Aiming at diminishing circuit complexity among others objectives, the form of these unitary operators can vary to suit the needs of different problems making the expansion form simplest. Without loss of generality, the unitary expansion form of this work takes the form of Pauli words. The premise for our algorithm to have an advantage is that the quantum circuit complexity which is related to the number of expansion terms does not increase with the same power as the operator's exponentiation, and it is possible to give a pre-estimation of the number of expansion terms.

We can encode each Pauli word for a $n$-qubit system as a $2n$-dimensional vector by decomposing the operators and using a 0/1 representation, as illustrated in the following example. For a four-qubit system, each of its expansion term is a tensor product of four Pauli matrices. Take $X_1 Y_2 Z_3 I_4$ as an example, where the subscripts indicating the qubit each Pauli matrix acts on. $Y$ can be decomposed into $X \cdot Z$ multiplied by a constant, so $X_1 Y_2 Z_3 I_4$ can be decomposed into $[X_1 \otimes X_2\otimes I_3 \otimes I_4]\cdot[I_1\otimes Z_2 \otimes Z_3 \otimes I_4]$ multiplied by a constant, the left square bracket contains only Pauli X and the identity matrix, while the right square bracket contains only Pauli Z and the identity matrix. Next we encode $X/Z$ as 1 and $I$ as 0, then the original Pauli product term can be represented as an eight-dimensional vector $[11000110]$. The encoding is a map from $n$-qubit Pauli group $\mathcal{P}_n$ to linear space $\mathbb{F}_2^{2n}$: $$c \cdot X_1^{x_1} \dots X_n^{x_n} Z_1^{z_1} \dots Z_n^{z_n} \mapsto (\mathbf{x},\mathbf{z}),$$ where $c\in \{1, i, -1,-i\}$ and $\mathbb{F}_2^{2n}$ is $2n$-dimensional linear space on binary field. This map is a group homomorphism between $(\mathcal{P}_n, \cdot)$ and $(\mathbb{F}_2^{2n},+)$.
We refer to the aforementioned process as the vector encoding method for Pauli words. 

 For a $n$-qubit system, corresponding to a $2^n$-dimension Hilbert space, the set of Pauli words obtained by the Pauli expansion of any operator acting on it can form a group $G$ with $4^n$ elements. For a specific operator $A$, the set of Pauli words obtained by expanding any power of $A$ forms a subgroup $G'$ of $G$. The original set of Pauli words for this operator $A$, denoted as $S$, serves as the generating set of the subgroup $G'$.
 When we represent the elements in $S$ using the vector encoding method described above, if the resulting set of vectors $V$ contains $l$ linearly independent vectors, meaning these $l$ vectors can generate $V$ through the addition rule in $\mathbb{F}_2^{2n}$. By employing the homomorphism described in the preceding paragraph, $G'$ can be mapped to an $l$-dimensional subspace within $\mathbb{F}_2^{2n}$.
Consequently, the expansion of any power of the operator $A$ can have a maximum of $2^l$ terms. And we hope that the value of $l$ should be as small as possible in order for improving the efficiency of our algorithm. For condensed systems, because of their periodicity and lattice symmetry, their Hamiltonians should also have a higher symmetry structure, and we suppose that the number of unitary Pauli matrices of power of their qubit Hamiltonians form may thus also converge faster \cite{kummer1981construction,wigner1937consequences,lax2001symmetry}.
In Appendix \ref{numberofpauliwords}, we have selected four models and presented the relationship between the exponentiation of their Hamiltonians and the number of terms in the Pauli expansion. 

\subsection{Quantum algorithm realization}

For quantum chemistry and the band structure problems concerned, we generally start from a linear combination form of Pauli words as equation \ref{Hlcu}. When determining the band structure of a crystal, given a certain wave vector $\vec{k}$, we can express the Hamiltonian as $H_{k}= \sum_{i=0}^{M_{k}-1} \alpha_{i}^{(k)} P_i$ where $M_{k}$ is the number of Pauli words and $m_{k} = \left \lceil log_2 M_{k} \right \rceil$. Let the set of eigenvalues be $\left \{ E_{j,k} \right \}^{2^n}_{j=1}$ with corresponding eigenstates $\left \{ \varphi_{j,k} \right \}_{j=1}^{2^n}$. In general crystal systems, there are always some bands with eigenvalues greater than zero. In order to ensure that all eigenvalues are less than zero, a bias term needs to be introduced and the Hamiltonian is  reconstructed as
\begin{equation}
    U_{1,k}= H_{k}-\lambda_0 I^{\otimes n}(\lambda _0>\max\left \{ 0,E_{1,k},\cdots E_{2^n,k}\right \}).
\end{equation}

Typically, the larger the bias parameter $\lambda_0$ is, the more time-consuming the algorithm becomes. Therefore, it is desirable to minimize the value \cite{wen2021full}. Assuming that the value of power required to obtain the ground state for $U_{1,k}$ is denoted as $t_{1,k}$, then we can realize the exponentiation
\begin{equation}
    U_{1,k}^{t_{1,k}}= \sum^{L_{1,k}-1}_{i=0} \beta_{i}^{(1,k)}P_i
\end{equation}
in the quantum circuit with $L_{1,k} \le 4^n$. As Appendix \ref{numberofpauliwords} shows, in many practical physical systems, it is often observed that $L_{1,k}$ is significantly smaller than $4^n$, and converge faster towards a relatively small value as $t_{1,k}$ increases. The coefficients $\beta_i^{(1,k)}$ of each Pauli word can be obtained through simple classical calculations, and we now shift our focus to the part of quantum circuit implementation of P-FQE. 

The first step involves initializing the ancillary and work systems.
Using the computational basis $\left |i\right \rangle_s$,
$\left |0\right \rangle_s$ refers to $\left |0\right \rangle^{\otimes l_{1,k}}$,
where $l_{1,k}=\left \lceil log_2 L_{1,k} \right \rceil$ is the number of ancillary qubits.
The ancillary qubits are introduced to create a larger Hilbert space,
and are initialized from $\left |0\right \rangle_s$ to a specific state
\begin{equation}
    \left | \psi_{1,k} \right \rangle = \frac{1}{\mathfrak{C}} \sum_{i=0}^{2^{l_{1,k}}-1} \beta_i^{(1,k)} \left |i \right \rangle_s,
\end{equation}
where $\mathfrak{C}=\sqrt{\sum_{i=0}^{2^{l_{1,k}}-1} |\beta_1^{(1,k)}|^2} $ is a normalization constant.
And the order of Pauli word is unimportant, let $\beta_i^{(1,k)}=0$ for $L_{1,k}\le i\le 2^{l_{1,k}}-1$ for simplicity.
An appropriate basis set such as linear combination of atomic orbitals (LCAO)
and orthogonalized plane wave (OPW) selected to initialize work system can make our algorithm more efficient \cite{ozaki2003variationally,phillips1959new}, and denote the trial initial state prepared as $\left | \varPhi_{0}^{(1,k)} \right \rangle$. In complex systems, the quantum random access memory (qRAM) method can be utilized to prepare the initial states \cite{giovannetti2008quantum}.

The second step of the quantum circuit involves entangling the ancillary qubits with the work qubits using a series of controlled gate operations. If the ancillary state is represented as $\left | i \right \rangle_s$, then the corresponding controlled gate acting on the work qubits would be Pauli word $P_i$. After entanglement, the entire system evolves into 
$(\sum_{i=0}^{L_{1,k}-1} \beta_i^{(1,k)} \left |i \right \rangle_s P_i \left | \varPhi_0^{(1,k)} \right \rangle) / \mathfrak{C}$ states.

The final step involves performing wave combination and measurement. By applying a Hadamard gate to each ancillary qubit, we execute the wave combination and the entire space is transformed into
\begin{equation}
    \frac{1}{\mathfrak{C}\sqrt{2^{l_{1,k}}}}(\left | 0 \right \rangle_s \otimes \sum_{i=0}^{L_{1,k}-1} \beta_i^{(1,k)} P_i \left | \varPhi_0^{1,k} \right \rangle + \sum_{i=1}^{2^{l_{1,k}}}\left | i \right \rangle_s \otimes \left | \Psi_i \right \rangle ),
\end{equation}
where $\left | \Psi_i \right \rangle$ is the state of work system in the subspace and the state of ancillary system is $\left | i \right \rangle_s$.
The work system collapse to the approximate ground state $|\tilde{\varphi}_{1,k}\rangle$ of $H_{k}$ when the ancillary system be measured as $\left | 0 \right \rangle_s$.

The probability of successfully obtaining the target state is $P_s = \left \| U^{t_{1,k}}_{1,k}\left | \varPhi_0^{(1,k)} \right \rangle \right \| /(\mathfrak{C}^2 2^{l_{1,k}} )$. While in the FQE algorithm, the probability of success is $P_{s(FQE)}=[\left \| U_{1,k}\left | \varPhi_0^{(1,k)} \right \rangle \right \| /(\mathbb{C}^2 2^{m_{k}})]^{T_{1,k}}$
declines exponentially with the number of iteration steps $T_{1,k}$ ($=t_{1,k}$ here) with $\mathbb{C} = \sqrt{\sum_{i=0}^{M_{k}-1} \alpha _i^2}$ \cite{rebentrost2019quantum,wei2020full}. And the noise in each iteration will further affect the probability of success. As long as $L_{1,k}$ does not exhibit exponential growth with respect to $M{k}$, we can conclude that our algorithm exponentially increases the probability of success compared to FQE, or equivalently, exponentially reduces the measurement complexity.

\begin{figure}[H]
    \centering{\includegraphics[width=\linewidth]{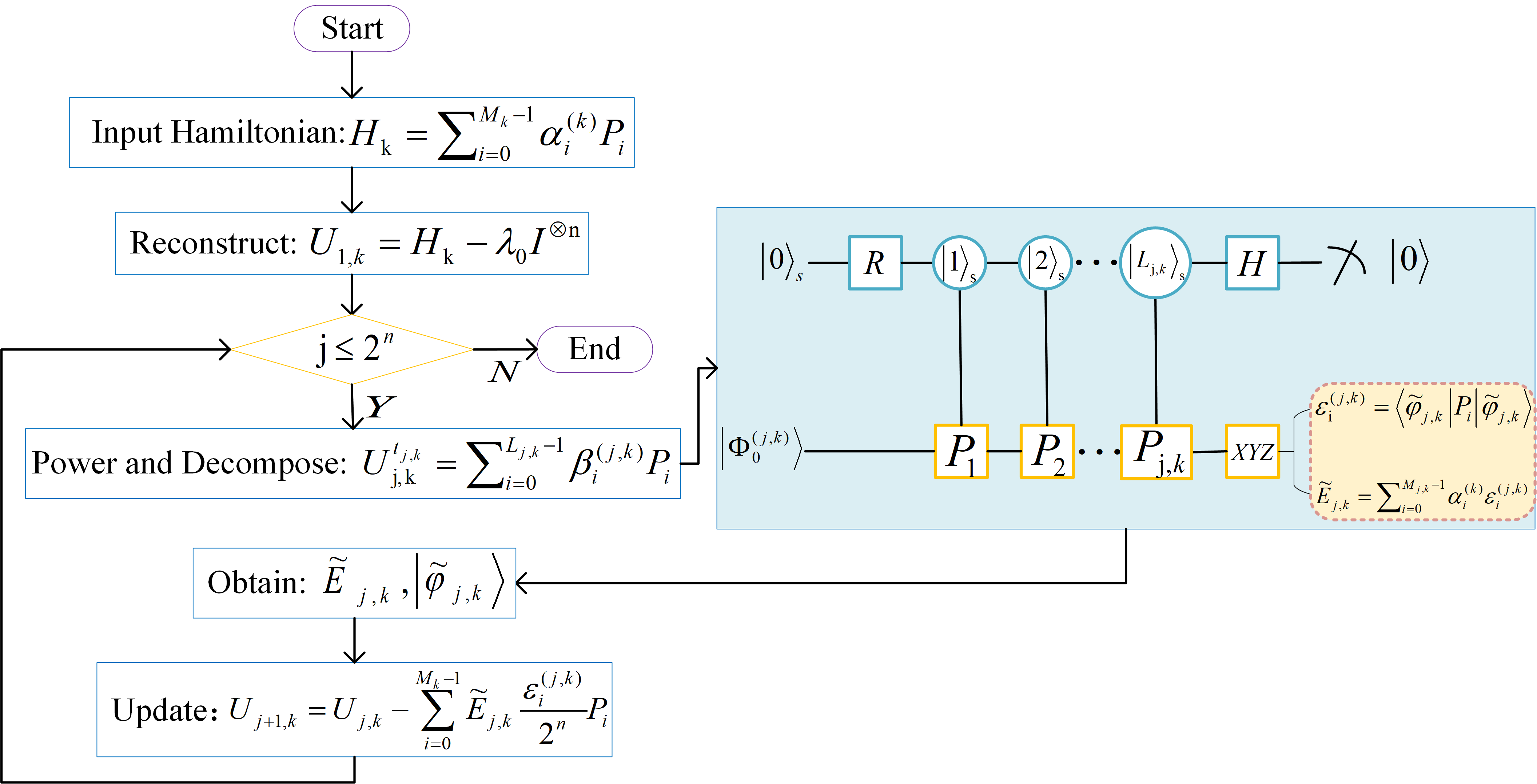}}
    \caption{The flow chart of P-FQE. The right panel illustrates the quantum circuit designed in our method.}
    \label{fig:gate_dec}
\end{figure}

We use Pauli measurements to estimate the approximate eigenvalue $\tilde{E}_{1,k}$ corresponding to $|\tilde{\varphi}_{1,k}\rangle$
\begin{equation}
    \varepsilon^{(1,k)}_i=\left \langle \tilde{\varphi}_{1,k}|P_i| \tilde{\varphi}_{1,k}  \right \rangle,
\end{equation}
\begin{equation}
    \tilde{E}_{1,k} = \left \langle \tilde{\varphi}_{1,k}  |H_{k}| \tilde{\varphi}_{1,k}  \right \rangle
=\sum_{i=0}^{M_{1,k}-1}\alpha_{i}^{(k)}\varepsilon^{(1,k)}_i.
\end{equation}
where $\varepsilon_i^{(1,k)}$ can be obtained by repeated measurements of Pauli word $P_i$. For obtaining the higher excited-state and excited-energy of $H_{1,k}$, we use a procedure similar to FQESS.
Update the original operator as
\begin{equation}
    \begin{split}
        U_{2,k}&= H_{k}-\tilde{E}_{1,k}|\tilde{\varphi}_{1,k}\rangle \langle\tilde{\varphi}_{1,k}|-\lambda _0 I^{\otimes n}\\
        &=\sum_{i=0}^{M_{k}-1} (\alpha_{i}^{(k)}-\tilde{E}_{1,k}\frac{\varepsilon_{i}^{(1,k)}}{2^n})P_i-\lambda_0 I^{\otimes n}\\
       &=\beta_i^{(2,k)}P_i ,
        \end{split}
\end{equation}
of which the ground state is the second excited state of $H_k$.
And the new operator implemented in the quantum circuit is $U_{2,k}^{t_{2,k}}=\sum_{i=0}^{L_{2,k}-1}\beta_i^{(2,k)}P_i$.
The subsequent steps follow the same procedure as described above.
But in order to obtain the energy spectrum corresponding a specific $k$-point, it may be necessary to initialize the work qubits in mutually orthogonal states for different energy levels, thereby ensuring the
orthogonality between the eigenstates associated with different energy levels. Iterating the aforementioned procedure, the whole energy spectra for Hamiltonian $H_k$ can be obtained.
By systematically varying the values of $k$ along high-symmetry paths and obtaining the corresponding Hamiltonians, we can repeatedly perform the previously described procedure to determine the complete energy band structure of the target crystal. The detailed steps of the P-FQE algorithm are illustrated in the algorithm flowchart, as shown in FIG. \ref{fig:gate_dec}.

\subsection{Complexity analysis}
The complexity of P-FQE algorithm consists of classical and quantum computation parts. We take the calculation of energy spectrum $H_k$ as an example.
The Pauli expansion of the operator calculated in classical computer is equivalent to the multiplication of weight coefficient of each Pauli word,
and the weight coefficients below a certain threshold are discarded during the process. The complexity of classical part is $O(M_k^{t_k})$, where $t_k = \max\left \{t_{j,k} \right \}$.
For the complexity of quantum computing part, we consider gate complexity and measurement complexity. The maximum total basic gates needed to obtain some energy level is $O(nL_k\log_2L_k)$, where $L_k=\max\left \{ L_{i,k}\right \}$ \cite{wen2021full,long2001efficient,xin2017quantum,wei2018efficient}.
The measurement complexity is $O(\sqrt{L})$ when the technique of quantum amplitude amplification is employed to further enhance the probability of success \cite{brassard2002quantum,berry2015simulating}. 

\section{Numerical Simulation}
\label{sec:numericalsimulation}
\begin{figure}
    \centering{\includegraphics[width=0.9\linewidth]{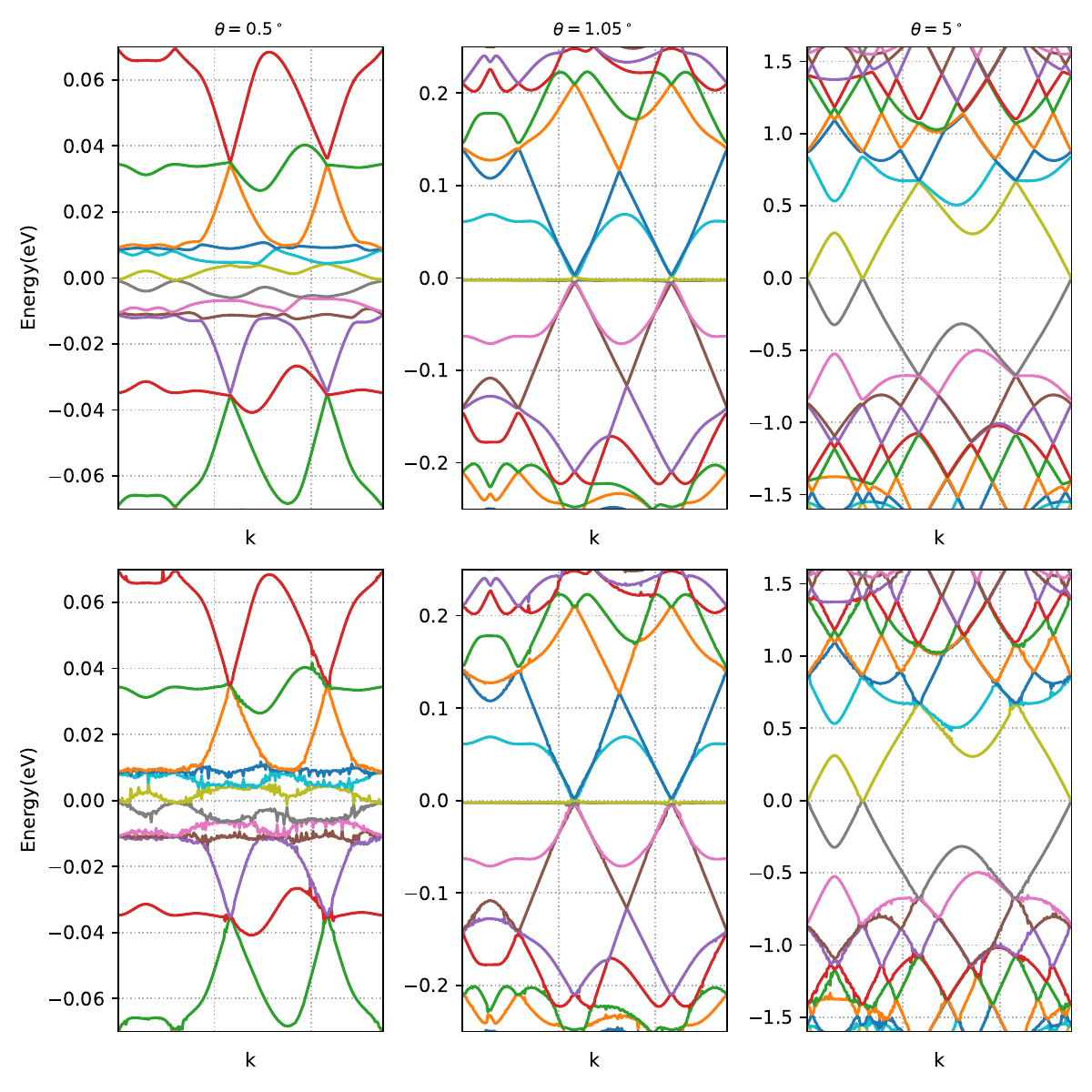}}
    \caption{The comparison of Moiré bands in twisted double-layer graphene obtained from classical diagonalization (the first row) and numerical
    simulation of our P-FQE algorithm (the second row). Here $w = 110~meV$, and twist angles $\theta = 0.5^{\circ}$ (the first column), $1.05^{\circ}$ (the second column), and $5^{\circ }$ (the third column).}
    \label{im2}
\end{figure}

We use twisted double-layer graphene as the demonstration model to numerically validate our algorithm. In bilayer graphene, the electronic band structure undergoes changes due to interlayer interactions, leading to tunable electronic properties. Moreover, different stacking structures can be achieved by adjusting the twist angle between the two layers. And the composite structure will give rise to a set of long-period patterns known as Moiré patterns
on the underlying atomic lattice for certain twist angles \cite{bistritzer2011moire,aspuru2005simulated,oster1964theoretical}. Bistritizer and MacDonald~\cite{bistritzer2011moire} constructed a low-energy continuum effective Hamiltonian for twisted bilayer graphene in a tight-binding representation that is applicable to arbitrary translation vector and twist angles $\theta \preceq 10 ^{\circ}$, independent of whether or not the structure is commensurate. 
The Bistritizer-MacDonald model is as follows
\begin{equation}
    H^k(\mathbf{r})=\begin{pmatrix}
        -i\nu_F\mathbf{\sigma}\cdot\nabla  & T(\mathbf{r})\\
         T^{\dagger}(\mathbf{r}) &-i\nu_F\mathbf{\sigma}\cdot\nabla
      \end{pmatrix}.
\end{equation}
where $T(\mathbf{r})$ is the interlayer hopping term, and $\nu_F$ is the Fermi velocity. We set the parameters to be the same as those in paper \cite{bistritzer2011moire} with lattice constant $a_0 = \sqrt{3} \cdot 1.42 ~\AA$ and interlayer hopping energy $w = 110~meV$. The plane wave expansion method is used to solve numerically for the electronic spectrum and truncating the momentum-space lattice at the third honeycomb shell. The size of the Hamiltonian matrix is 196$\times$196. The Moiré bands structure for three twisted angles $\theta= 5^{\circ}$, $1.05^{\circ}$, and $0.5^{\circ }$ obtained through numerical simulation of P-FQE and matrix diagonalization method is shown in FIG. \ref{im2}. For twist angles of  $5^{\circ}$, $1.05^{\circ}$, and $0.5^{\circ }$, the bias term $\lambda_0$ were set to 9.8, 2.3, and 1.2, respectively. And we select a set of randomly orthogonal basis states as the initial states for the work qubits at each $k$-vector when computing different energy bands. For convenience, we directly use the 400th power of the initial Hamiltonian as the operator acting on the initial state in the quantum circuit for $\theta = 0.5^\circ$ and $1.05^\circ$, and 230th power for $\theta = 5^\circ$.

The Moiré bands in the first row represent the theoretical results obtained by classical processing, which are consistent with those presented in paper \cite{bistritzer2011moire}. The Moiré bands in the second row are the numerical simulations obtained using our algorithm, which are in good agreement with the theoretical expectations. The number of bands increases within a specific energy range and band separation decreases as the twist angle decreases, leading to a higher density of states. When $\theta=1.05^{\circ}$, Dirac-point velocity vanishes and a very flat band appears which implies strong correlation. It is worth mentioning that the spectrum results at smaller energy-gap points will be more sensitive to the power values, which are not as smooth as other points under same parameter setup. It can be improved by improving the order of power or introducing suitable bias values.

\section{Results on Real Quantum Devices}\label{sec:results}
In this part, we present two experimental demonstrations of the P-FQE algorithm with different physical systems (Weyl semimetal and graphene) on the superconducting quantum computing chips from Quafu/IBM quantum cloud platforms. Details of the two chips can be found in Appendix \ref{cloudplatform}, and we introduce the experimental process and results below.

\subsection{Weyl semimetal}
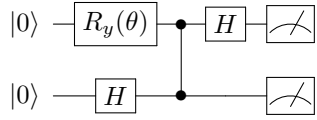
\begin{figure}[H]
    \centerline{
      \Qcircuit @C=.8em @R=1.2em {
      \lstick{\ket{0}} &\gate{R_y(\theta)} &\ctrl{1}& \gate{H} & \meter \\
      \lstick{\ket{0}} &\gate{H} &\control \qw &\qw &\meter
      }
      }
      \caption{Quantum circuit for obtaining energy bands for the minimal model of a Weyl semimetal. Only one ancillary qubit and one work qubit are needed, with a controlled-Z gate required to entangle the two qubits. The rotation angle $\theta$ in $R_y(\theta)$ gate is modified as a function of energy levels and wavevector $k_z$ in the simulation.} \label{im3}
\end{figure}

The first model we chose is Weyl semimetal, a special topological material with nontrivial band structure. In Weyl semimetals, the valence band and conduction band intersect at specific momentum space positions, forming Weyl points \cite{armitage2018weyl,wang20173d,zhang2016linear,okugawa2014dispersion}. 
The energy bands near these points exhibit Dirac-like linear dispersion with relativistic features, forming a cone-shaped Fermi surface known as the Weyl cone.
Weyl semimetals are of significant research interest in the field of topological physics and hold great potential for applications in novel electronic devices, optical materials, and topological quantum computation. The Hamiltonian for the minimal model for a Weyl semimetal is \cite{lu2015high}
\begin{equation}\label{eq14}
    H = A(k_x*\sigma_x+k_y * \sigma_y)+[M_0-M_1(k_x^2+k_y^2+k_z^2)]\sigma_z 
\end{equation}
where parameters are set as $M_0=M_1=A=1,k_x=k_y=0$. So the Hamiltonian can be reduced to $H_{kz}=(1-k_z^2)\sigma_z$, and the energy spectrum is a function of $k_z$. We set the bias parameter $\lambda_0 = 4$, and the value of power $t_{j,k}=20$. Then the operator implemented on the quantum circuit is $(H_{k_z}-4I)^{20}$ for ground energy band. After obtaining the the ground energy $E_{1,k_z}$ and corresponding ground state $|\varphi_{1,k_z}\rangle$ for $H_{k_z}$,
we implemented $(H_{k_z}-E_{1,k_z}|\varphi_{1,k_z}\rangle \langle\varphi_{1,k_z}|-4I)^{20}$ on the circuit
for obtaining the excited energy band.
This is a simple model that can be done with high fidelity on the quantum devices,
with one qubit as ancillary qubit and another one as work qubit.
Due to the peculiar nature of the eigenstates of the model, a Hadamard gate is applied to
the working qubit to initialize it in a superposition state of $\left | 0  \right \rangle$ and $\left | 1 \right \rangle$. The quantum circuit is as FIG. \ref{im3}.

\begin{figure}[H]
\centering{\includegraphics[width=\linewidth]{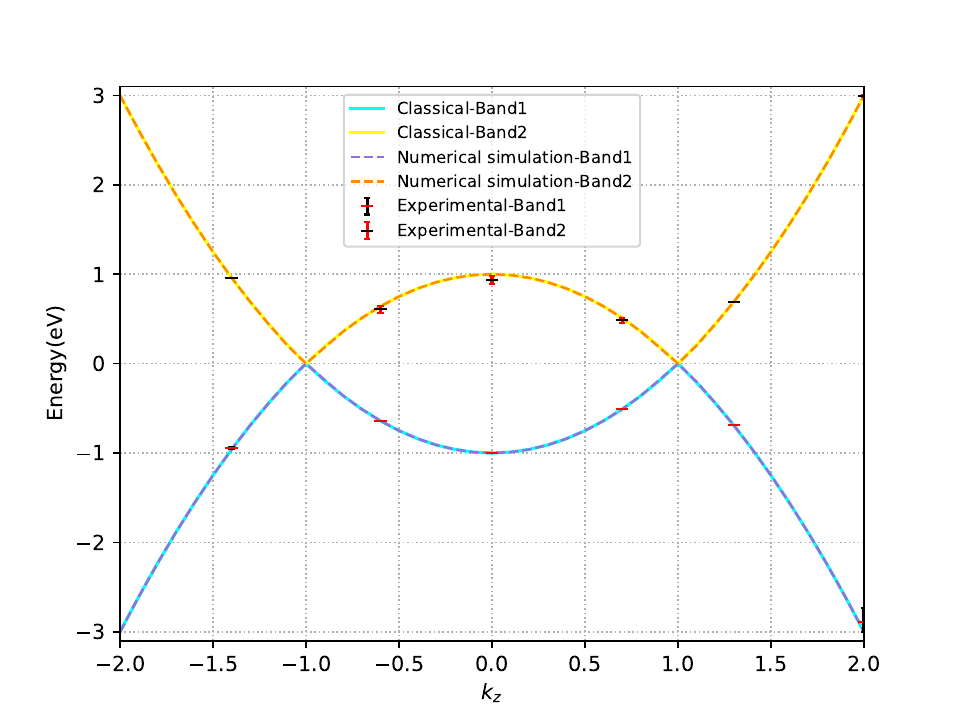}}
\caption{Experimental results of energy band structure for a Weyl semimetal as a function of $k_z$. The solid lines represent the results from classical computation, and the dashed lines represent the results from classical simulation of P-FQE algorithm. The error bars represent the experimental results on the superconducting quantum computing platform.}
\label{im4}
\end{figure}

FIG.\ref{im4} presents band spectrum of a Weyl semimetal obtained from classical calculation (sold lines), numerical simulation of P-FQE (dashed line), and experimental results of P-FQE algorithm (error bar). The experiments are conducted on the Quafu quantum cloud platform \cite{quafu}. We take seven values of $k_{z}\in \{-2, -1.4, -0.6, 0, 0.7, 1.3, 2\}$ in the experiment. For different $k_z$ and energy bands, the Hamiltonian would change and the auxiliary qubit would rotate at a different angle around the Y-axis accordingly, with an unchanged form of the quantum circuit. For each $k$-vector, we performed three experimental trials and taken 40,000 samples in each trial. The error bars plotted using the average, minimum, and maximum values of the three trials. We can see that the experimental results of P-FQE algorithm are in good agreement with the theoretical values and numerical simulation. Two energy bands intersect at symmetric points, forming a pair of Weyl points. Due to the simplicity of the model, the eigenstates can often be measured accurately in experiments, resulting in very small or even zero error bar ranges.

\subsection{Graphene}
The second model for the experiments is single-layer graphene \cite{geim2009graphene,li2016graphene,abergel2010properties,konschuh2010tight}, a two-dimensional crystal with a hexagonal honeycomb lattice structure composed of single-layer carbon atoms exhibiting SP2 hybridization.
This distinctive lattice arrangement leads to the conduction and valence bands of graphene intersecting at the $K$ point in the Brillouin zone, resulting in its zero-bandgap semiconductor characteristics. The Hamiltonian has a form of 
\begin{equation}\label{eq15}
\begin{split}
        H_{k} = &(2\cos\frac{k_y}{2\sqrt{3}}\cos\frac{k_x}{2}+\cos\frac{k_y}{\sqrt{3}})\sigma_x\\
        +&(2\sin\frac{k_y}{2\sqrt{3}}\cos\frac{k_x}{2}-\sin\frac{k_2}{\sqrt{3}})\sigma_y    
\end{split}
\end{equation}
and there are three terms ($I$, $\sigma_x$, and $\sigma_y$) after introducing a bias term $-4I$.
Because $(A\sigma_x+B\sigma_y-4I)^n=a\sigma_x +b\sigma_y+cI$ with $n=25$ here, we just need two ancillary qubits and one work qubit. 
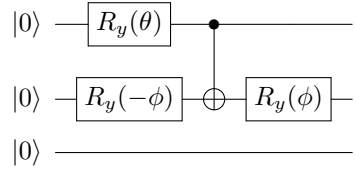
\begin{figure}[H]
    \centerline{
      \Qcircuit @C=.8em @R=1.2em {
      \lstick{\ket{0}} &\gate{R_y(\theta)} &\ctrl{1} & \qw & \qw \\
      \lstick{\ket{0}} &\gate{R_y(-\phi)} &\targ & \gate{R_y(\phi)} & \qw\\
      \lstick{\ket{0}} &\qw &\qw &\qw &\qw
      }
      }
      \caption{The initialization part of the quantum circuits for calculating the band structure of single-layer graphene. The ancillary qubits be initialized to the state of $\cos \frac{\theta}{2}\ket{0}\ket{0}-\sin\phi \sin\frac{\theta}{2}\ket{1}\ket{0}+\cos\phi \sin\frac{\theta}{2}\ket{1}\ket{1}$.} \label{im5}
\end{figure}

\begin{figure}[H]
    \centering{\includegraphics[width=7.5cm,height=2.5cm]{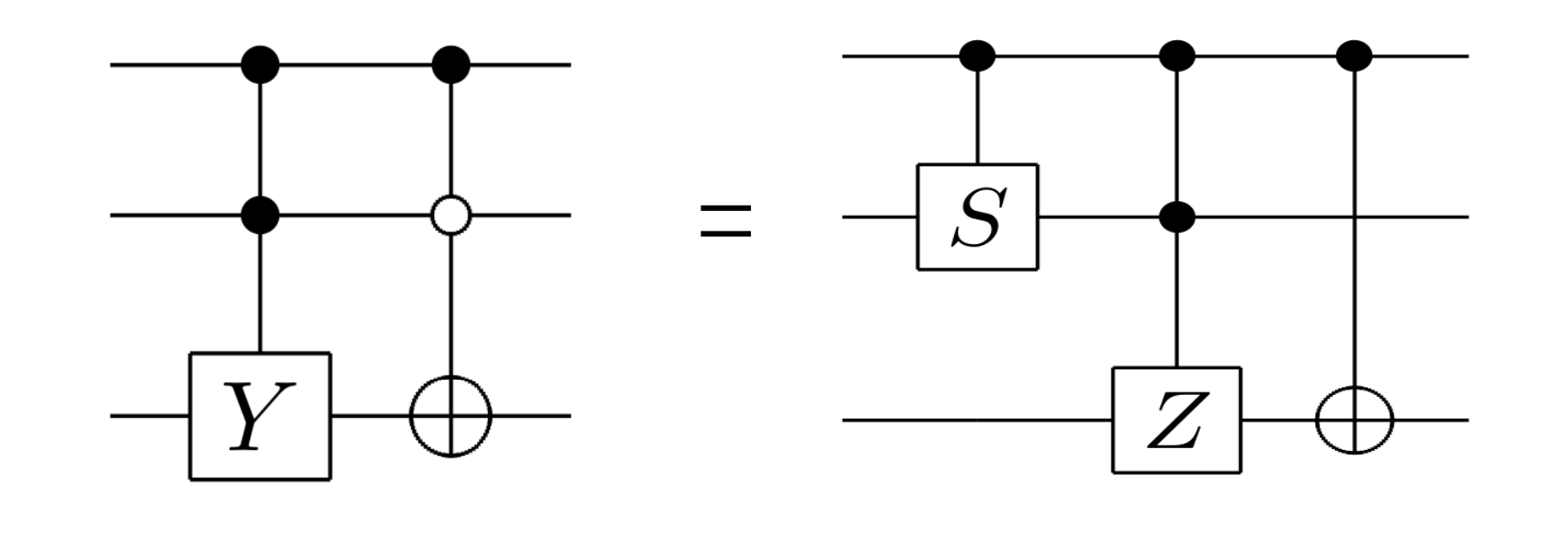}}
    \caption{The left $C^2(Y)-COX$ gate is equivalent to the right circuit. } \label{im6}
\end{figure}

\begin{figure}[H]
    \centerline{
      \Qcircuit @C=.8em @R=1.2em {
      \lstick{\ket{0}} &\gate{R_y(\theta)} &\ctrl{1}& \qw & \ctrl{1} & \qw & \ctrl{1}&\qw&\qw&\ctrl{1}&\ctrl{2}&\ctrl{2}&\gate{H} &\meter\qw \\
      \lstick{\ket{0}} &\gate{R_y(-\phi)} & \targ &\gate{R_y(\phi)} & \gate{S}&\ctrl{1}&\targ &\ctrl{1}&\ctrl{1}&\targ&\qw&\qw&\gate{H} &\meter\qw\\
      \lstick{\ket{0}} & \qw & \qw & \qw & \qw & \gate{S} & \qw & \gate{Z} & \gate{S} & \qw & \gate{S} & \targ & \gate{H} & \meter\qw
      }
      }
      \caption{Quantum circuit for computing the band structure of single-layer graphene on superconducting quantum computer IBM-nairobi. The figure displays the circuit diagram for computing the first band and measuring the Pauli word X. When computing the second energy band, add a X gate to initialize the work qubit to the state $|1\rangle$. When measuring the Pauli word Y, replace the Hadamard gate acting on the work qubit before measurement with a rotation gate $R_x(\pi/2)$.} \label{im7}
\end{figure}
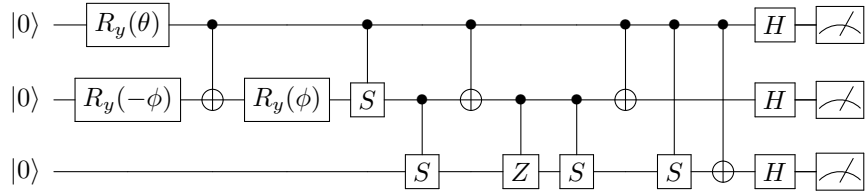

The circuit can be divided into three parts. The first part is to initialize the ancillary register from $\ket{0}\ket{0}$ to the superposition state $\cos \frac{\theta}{2}\ket{0}\ket{0}-\sin\phi \sin\frac{\theta}{2}\ket{1}\ket{0}+\cos\phi \sin\frac{\theta}{2}\ket{1}\ket{1}$,
which correspond respectively to term $I$, term $\sigma_x$ and term $\sigma_y$, as shown in FIG. \ref{im5}. The second part is controlled operations implemented on the work qubit (COX and CCY). Because of $\sigma_y =i\sigma_x \sigma_z$, a simplification of circuit can be done as FIG. \ref{im6}. The $C^2(Z)$ can be further decomposed to a series of C-U gates. The third part is measurement. We only need measure $\sigma_{x}$ and $\sigma_{y}$ results by applying Hadamard gate and $R_x(\pi/2)$ gates before measuring in the computational basis. The work qubit would be initialized to $\ket{1}$ by applying X gate for calculating the second energy band. The final quantum circuit we run on the IBM cloud platform is shown in FIG. \ref{im7}, which applied for the first band and measure Pauli-X.

\begin{figure}[H]
    \centering{\includegraphics[width=\linewidth]{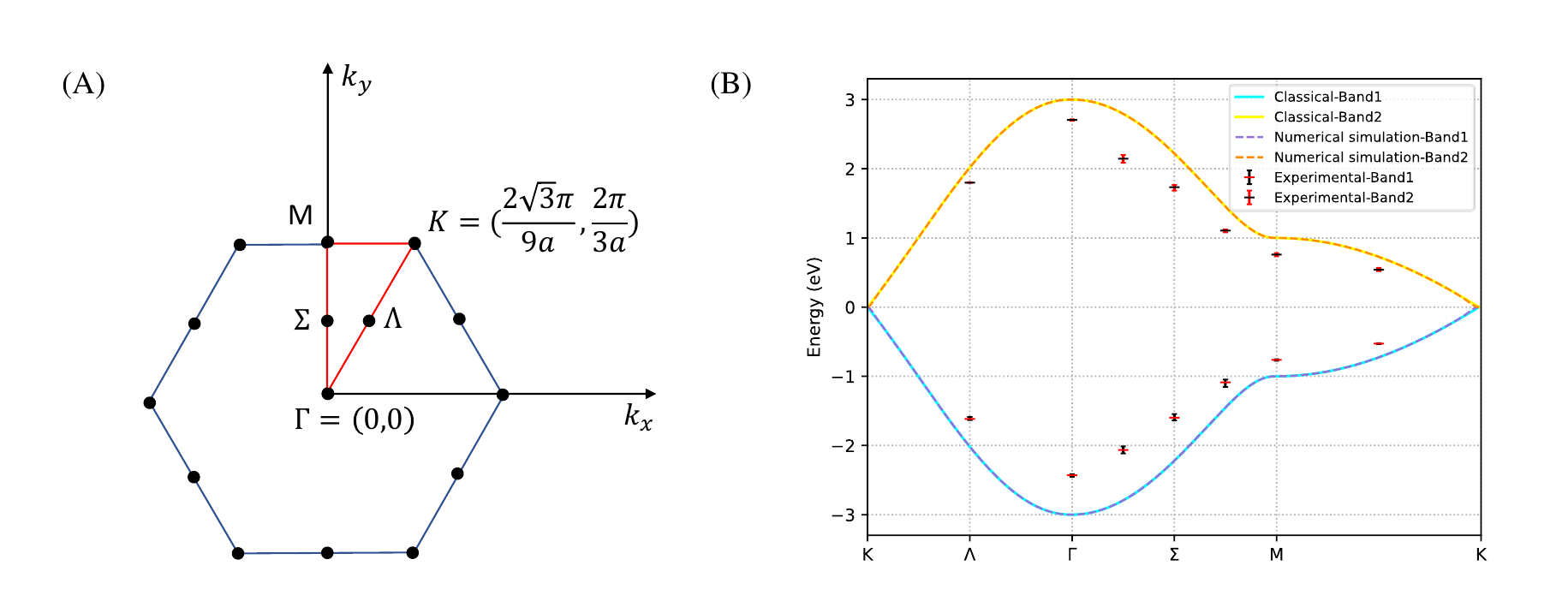}}
    \caption{The energy band structure of single-layer graphene. (a) The Brillouin zone of single-layer graphene with the high-symmetry $k$-points with $\Lambda=(\frac{\sqrt{3}\pi}{9a}, \frac{\pi}{3a})$, $\Gamma=(0, 0)$, $\Sigma=(0, \frac{\pi}{3a})$, $M=(0, \frac{2\pi}{3a})$. (b) The energy bands depicted along the the axis of high symmetry ($K\to \Gamma \to M\to K$) in the momentum space of graphene. Solid lines mark the energy bands obtained from classical computations, while dashed line represent the results obtained through numerical simulation of the P-FQE algorithm. The error bars illustrate the experimental measurements of the P-FQE algorithm conducted on IBM-nairobi chips.}
    \label{im8}
\end{figure}

Due to the relatively high complexity of the second experimental model, requiring a deeper quantum circuit and a larger number of control gates, it imposes higher hardware requirements. Therefore, we chose the IBM quantum chip---IBM-nairobi as the platform for our second experiment. We selected seven experimental points along the high-symmetry path ($K\to \Gamma \to M\to K$) in the reciprocal space of graphene. 
In addition to the selected k-points $\Lambda:(\frac{\sqrt{3}\pi}{9a}, \frac{\pi}{3a})$, $\Gamma:(0, 0)$, $\Sigma:(0, \frac{\pi}{3a})$, and $M:(0, \frac{2\pi}{3a})$, we have also chosen one k-point along each of the paths from $\Gamma$ to $\Sigma$, $\Sigma$ to $M$, and $M$ to $K$. These additional k-points are $(0, \frac{3\pi}{18a})$, $(0, \frac{\pi}{2a})$, and $(\frac{\sqrt{3}\pi}{9a}, \frac{2\pi}{3a})$. 
For the first two $k$-points, the absence of the Pauli Y term in their Hamiltonian allows us to perform the experiments using only two qubits in practice. For each $k$-point, we conducted three trials and plotted error bars, with each experiment consisting of 40,000 samples. From FIG. \ref{im8}, it can be observed that the numerical simulation results of P-FQE algorithm align perfectly with the theoretical values. However, there may be slight discrepancies in the experimental results, yet the overall trend remains consistent. In Appendix \ref{experimentaldata}, the experimental data of the two experiments are presented.
\section{Conclusions}\label{chap-con}

In summary,  P-FQE algorithm addressed the issue of exponential decay of success probability as the number of iterations in FQE increases, by substituting the original operator with its powers. Meanwhile, P-FQE does not require much additional ancillary qubits. We use this algorithm to study band structure calculations. Our algorithm trades classical computational complexity for the probability of successfully measuring the target state in the quantum computer. We have proposed the vector encoding method to estimate the maximum number of expansion terms of the power of operators, which can also be used to determine the order where a Taylor series can be truncated as in Ref\cite{berry2015simulating}. P-FQE algorithm is more suitable for current quantum computers because of reducing the real runs. We present a numerical simulation of the calculation of twisted double-layer graphene model, and two experiments on different superconducting computers to validate the effectiveness and feasibility of P-FQE algorithm. In future, we will try to find the optimal balance between the number of powers and the complexity of the algorithm to increase the efficiency of our algorithm. Furthermore, we will try to establish a more precise relationship between the exponentiation of Hamiltonians for various physical systems and the number of terms in their expansion using linear combination of unitary operators.

\section{Acknowledgements} 
This research was supported by National Basic Research Program of China. S.W. acknowledge the National Natural Science Foundation of China under Grants No. 12005015. We gratefully acknowledge support from the National Natural Science Foundation of China under Grants No. 11974205, and No. 11774197. The National Key Research and  Development Program of China (2017YFA0303700); The Key Research and  Development Program of Guangdong province (2018B030325002); Beijing Advanced Innovation Center for Future Chip (ICFC).

\appendix
\section{Error estimation and the order of power estimation}
\label{error estimation}

In subsection \ref{sec:iteration method}, Denote $\frac{A^k x^{(0)}}{\sqrt{x^{(0)\dagger}A^{2k}x^{(0)}}} = \left |x^{(k)} \right \rangle $, then the error $\epsilon$ can be estimated,

\begin{equation}
\begin{split}
    \epsilon & = \left \langle x^{(k)}\right | A \left | x^{(k)} \right \rangle -\lambda _0 \\
    &=\frac{\sum_{i=0}^{n-1}  a_i^2 \lambda _i^{2k+1}}{\sum_{i=0}^{n-1} a_i^2 \lambda _i^{2k}}-\lambda_0\\
    &=\frac{\sum_{i=1}^{n-1} a_i ^2\lambda _0^{2k} (\frac{\lambda _i}{\lambda _0})^{2k}(\lambda_i - \lambda _0)}{\sum_{i=0}^{n-1} a_i^2 \lambda _0^{2k}(\frac{\lambda _i}{\lambda _0})^{2k}} \\
    &=\frac{\sum_{i=1}^{n-1} a_i ^2 (\frac{\lambda _i}{\lambda _0})^{2k}(\lambda_i - \lambda _0)}{a_0^2 + \sum_{i=1}^{n-1} a_i^2(\frac{\lambda _i}{\lambda _0})^{2k}}\\
    &\le \frac{(n-1)a_m^2}{a_0^2}(\lambda _n - \lambda _0)(\frac{\lambda _1}{\lambda _0})^{2k}. 
\end{split}
\end{equation}
The error decreases exponentially with $2k$ (twice the order of the power). That means, if we aim to achieve a preset precision $\epsilon$, the value of the order of the power need be
\begin{equation}
    k = O(log\frac{n}{\epsilon }).
\end{equation}

\section{Relationship of terms number in Pauli expansion and order of the Hamiltonian power}
\label{numberofpauliwords}

\begin{figure}[H]
    \centering{\includegraphics[width=\linewidth]{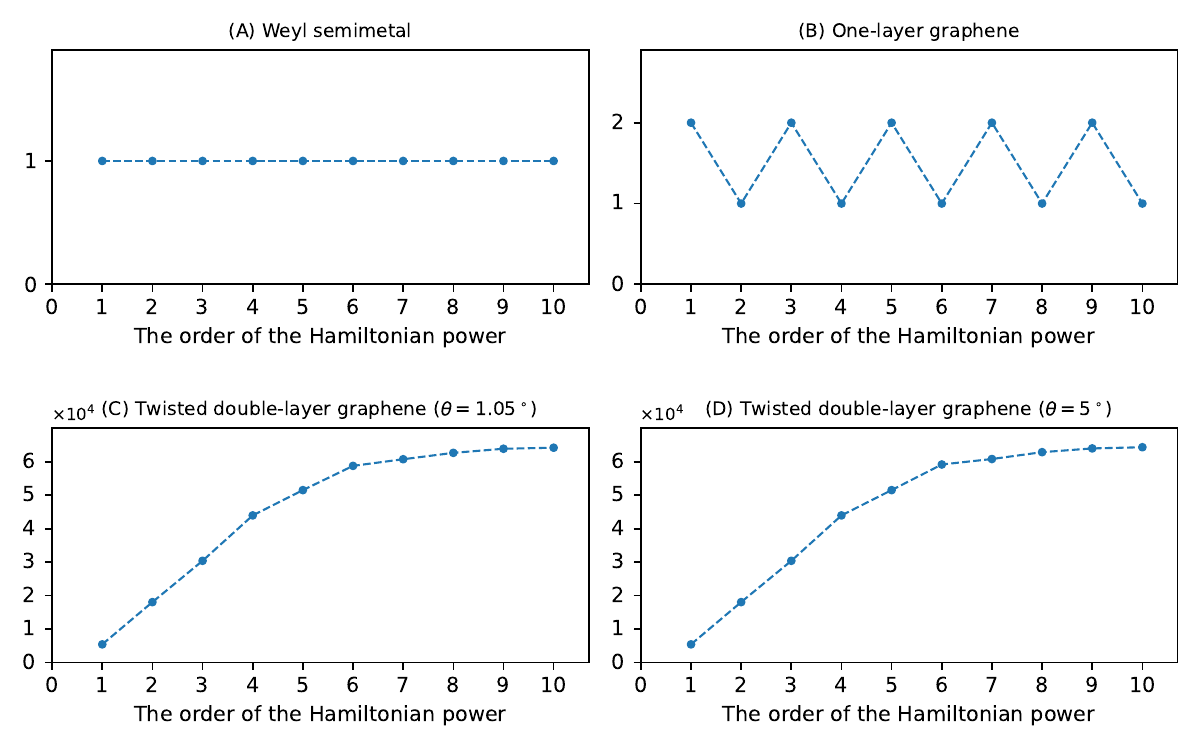}}
    \caption{The relationship of the number of terms in the Pauli expansion and the order of the Hamiltonian power. (A) using the minimal model for a Weyl semimetal in Eq. \ref{eq14} and $k_x=k_y=k_z=0$. (B) using the model of one-layer graphene in Eq. \ref{eq15} and $kx=ky=2\pi$. (C) and (D) using models of twisted bilayer graphene with $k_x=0$,$k_y=1$ for rotation angles $1.05^{\circ}$ and $5^{\circ}$, respectively. The size of their Hamiltonian matrix is $196\times196$, while for a Hermitian matrix of the same dimension, the number of terms in the Pauli expansion can be $4^8$ at most.}
    \label{im9}
\end{figure}

Fig. \ref{im9} illustrates the relationship between the number of terms in the Pauli expansion and the order of the Hamiltonian power for four different Hamiltonian models.
From the figure, we can observe three scenarios where, with increasing powers of the Hamiltonian, the number of terms in the Pauli expansion either converges rapidly, fluctuates, or the increase slowed rapidly and limited in a small range. These three scenarios often apply to other physical systems as well. 

\section{Quafu quantum cloud platform and IBM Quantum}
\label{cloudplatform}

Quafu is an open quantum computing cloud platform \cite{quafu}, connected to a 136-qubit quantum computer and two other 18 and 10-qubit superconducting quantum computers. In the experiment of Weyl semimetal, we use the first two qubits of the 18-qubit processor named ScQ-P18, with qubit frequency of 4.59 and 5.02 GHz, and has up to $97.3\%$ average fidelity of  two qubit logic gate Controlled-Z.

\begin{figure}[H]
    \centering{\includegraphics[width=\linewidth]{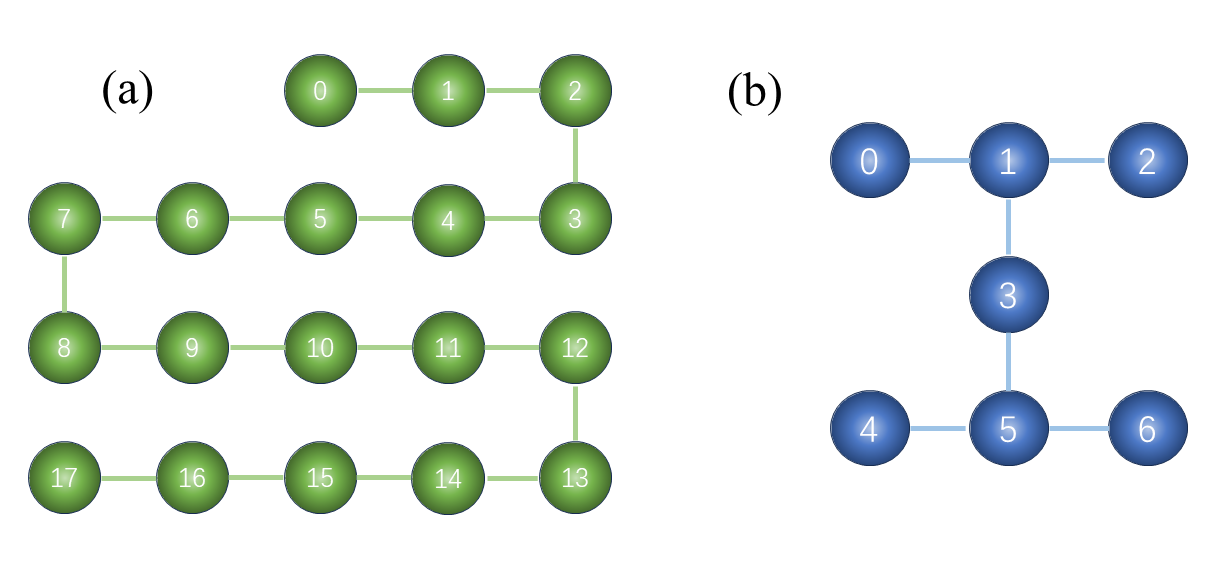}}
    \caption{The topological maps of superconducting quantum chips ScQ-P18 and $ibm_nairobi$. Each qubit coupled with its nearest-neighbors. (a) The topological map of backend ScQ-P18 on Quafu. (b) The topological map of backend $ibm_nairobi$ on IBM Quantum.}
    \label{im10}
\end{figure}

In the experiment of computing the energy band of graphene, we chose the quantum backend $ibm_nairobi$ on IBM quantum cloud platform, a superconducting quantum chip with 7 qubits. The basis gates of $ibm_nairobi$ are CX, ID, RZ, SX, and X, and the median errors for the CNOT gate, SX gate, and readout are $9.360\times10^{-3}$, $2.894\times10^{-4}$, and $2.500\times10^{-2}$. The first three qubits are used in our circuits, with work frequency of 5.26, 5.17 and 5.274 GHz. The topological maps of ScQ-P18 and $ibm_nairobi$ are shown in the FIG. \ref{im10}.

\section{Experimental data}
\label{experimentaldata}

Table \ref{table1} and Table \ref{table2} are the results of experiments on two superconducting quantum computers. The error is calculated by $(Average\; value-Theoretical\;value)/Theoretical\;value$, where the $Average\;value$ is the average of three trial results.

\begin{table}[H]
    \centering
    \caption{The experimental results of Weyl semimetal on the QuaFu quantum cloud platform using a 10-qubit chip (ScQ-P10).}
    \label{table1}
    \begin{tabular}{cccccccc}
    \hline\hline
    K-point& Energy level&\multirow{2}{*}{Trial 1} & \multirow{2}{*}{Trial 2} & \multirow{2}{*}{Trial 3}& Average value &Theoretical value& Error \\

    $(k_x=k_y=0)$ &$n$&\multirow{2}{*}{}& \multirow{2}{*}{} & \multirow{2}{*}{} & $(Ev)$ &$(Ev)$& $(\%)$\\
    \hline

    \multirow{2}{*}{$k_z = -1.4$} & $n=1$& -0.9328 
    & -0.96 &-0.9398 &-0.9442& -0.96 & -1.645 \\

    \multirow{2}{*}{} & $n=2$& 0.96 & 0.96 & 0.96 & 0.96 & 0.96 & 0\\

    \multirow{2}{*}{$k_z = -0.6$} & $n=1$& -0.64 
    & -0.64 &-0.64 &-0.64& -0.64 & 0 \\

    \multirow{2}{*}{} & $n=2$& 0.5647 & 0.64 & 0.6211 & 0.6086 & 0.64 & -4.910\\

    \multirow{2}{*}{$k_z = 0$} & $n=1$& -1 
    & -1 &-1&-1& -1 &0\\

    \multirow{2}{*}{} & $n=2$& 0.8935 & 0.9159 & 0.9806 & 0.93 & 1 & -6.999\\

    \multirow{2}{*}{$k_z = 0.7$} & $n=1$& -0.51
    & -0.51 &-0.51 &-0.51& -0.51 & 0 \\

    \multirow{2}{*}{} & $n=2$& 0.4567 & 0.51 & 0.5041 & 0.4903 & 0.51& 3.872\\

    \multirow{2}{*}{$k_z = 1.3$} & $n=1$& -0.69
    & -0.6877 &-0.6821 &-0.6866& -0.69 & -0.491 \\

    \multirow{2}{*}{} & $n=2$& 0.69 & 0.69 & 0.69& 0.69& 0.69 & 0\\

    \multirow{2}{*}{$k_z = 2$} & $n=1$& -2.7315 
    & -3 &-2.9241 &-2.8852& -3 & -3.827 \\

    \multirow{2}{*}{} & $n=2$& 2.9787& 3& 3 &2.9929 & 3& 0.237\\
    \hline\hline

    \end{tabular}
    \label{experimentofweyl}
    \end{table}

    \begin{table}[H]
        \centering
        \caption{The experimental results of single-layer graphene on the IBM Quantum using the chip IBM-nairobi.}
        \label{table2}
        \begin{tabular}{cccccccc}
        \hline\hline
        K-point& Energy level&\multirow{2}{*}{Trial 1} & \multirow{2}{*}{Trial 2} & \multirow{2}{*}{Trial 3}& Average value &Theoretical value& Error\\
    
        $(k_x,k_y)$ &$n$&\multirow{2}{*}{}& \multirow{2}{*}{} & \multirow{2}{*}{} & $(Ev)$ &$(Ev)$& $(\%)$\\
        \hline
    
        \multirow{2}{*}{$(\frac{\sqrt{3}\pi}{9a}, \frac{\pi}{3a})$} & $n=1$&-1.5886 &-1.6339&	-1.6324&-1.6183&-2& -19.08
        \\
    
        \multirow{2}{*}{} & $n=2$& 1.7945 &	1.8008 &1.8040 &1.7998 &2&-10.010
        \\
    
        \multirow{2}{*}{$(0,0)$} & $n=1$& -2.4149&	-2.4513 &-2.4221 &-2.4294 &	-3	&-19.019
        \\
    
        \multirow{2}{*}{} & $n=2$& 2.6940 &2.7135 &2.7079 &2.7051 &3&-9.829
        \\
    
        \multirow{2}{*}{$(0, \frac{3\pi}{18a})$} & $n=1$& -2.0679&	-2.1165 &-2.0137 &-2.0660 &-2.7979&-26.158
        \\
    
        \multirow{2}{*}{} & $n=2$& 2.1502 &2.1970 &2.0870 &2.1447 &2.7979&-23.345
        \\

        \multirow{2}{*}{$(0, \frac{\pi}{3a})$} & $n=1$& -1.6400 &-1.5467 &-1.6052 &-1.5973 &-2.2361&-28.568
        \\
    
        \multirow{2}{*}{} & $n=2$& 1.6833 &1.7637 &1.7509 &1.7326 &2.2361&-22.515
        \\
    
        \multirow{2}{*}{$(0, \frac{\pi}{2a})$} & $n=1$& -1.0591 &-1.0491 &-1.1572 &-1.0885 &-1.4736&-26.136
        \\
    
        \multirow{2}{*}{} & $n=2$& 1.1220 &1.1112 &1.0810 &1.1047 &1.4763&-25.170
        \\
    
        \multirow{2}{*}{$(0, \frac{2\pi}{3a})$} & $n=1$& -0.7552 &-0.7755 &-0.7557 &-0.7621 &-1 &-23.786
        \\
    
        \multirow{2}{*}{} & $n=2$&0.7357 &0.7791 &0.7635 &0.7594 &1&-24.057
        \\

        \multirow{2}{*}{$(\frac{\sqrt{3}\pi}{9a}, \frac{2\pi}{3a})$} & $n=1$& -0.5303&	-0.5332&	-0.5230&	-0.5288&	-0.7321&-27.767
        \\
    
        \multirow{2}{*}{} & $n=2$&0.5662 &0.5308 &0.5206 &0.5392 &0.7321&-26.348
        \\
        \hline \hline
        
        \end{tabular}
        \label{experimentofgraphene}
        \end{table}

\bibliographystyle{apsrev4-1}
\bibliography{bibl.bib}

\end{document}